# Offsets of masers with respect to the middle of the Perseus arm and the corotation radius in the Milky Way

Jacques P. Vallée

National Research Council of Canada, Herzberg Astronomy & Astrophysics, 5071 West Saanich Road, Victoria, B.C., Canada V9E 2E7

## Abstract.

The radial distance to the corotation radius $R_{coro}$ (where the angular speed of the gas and stars $\Omega_{gas}$ in orbit around the Galactic Centre is equal to the angular speed of the spiral arm pattern $\Omega_{sp}$) has often been predicted (at various places), but not measured with a high precision.

Here we test the locations of masers with respect to the Perseus arm (Table 1). Our analysis of the masers and HII regions near the Perseus arm (mostly located on the *inner* arm side, by about 0.4 ±0.1 kpc from the cold CO mid-arm) shows that the corotation $R_{coro}$ must be > 10.8 kpc from the Galactic Center (Figure 1). This implies that the angular rotation speed of the spiral pattern $\Omega_{sp}$ < 21.3 km/s/kpc.

Another test in galactic quadrant II shows that the radial velocity of the masers are generally more negative than that of the CO mid-arm (Figure 2), indicating a deceleration with respect to the CO mid-arm, by about 9 ±3 km/s. This implies that $\Omega_{sp}$ < 20.7 km/s/kpc, and thus $R_{coro}$ > 11.1 kpc.

Finally, comparing our results with other published results (Table 2), we find a statistical mean corotation radius $R_{coro}$ predicted to be near 12 ±1 kpc from the Galactic Center (beyond the Perseus arm; before the Cygnus arm), and a mean angular spiral pattern speed $\Omega_{sp}$ predicted to be near 19 ±2 km/s/kpc.

## 1. Introduction

Tracer offsets (location of dust/masers versus cold molecular arm) over a whole galactic disk have been predicted by the density wave theory. Alternate theories also do not deny the existence of tracer offsets, but limit them to localized regions.

Assuming that the spiral arms are produced mainly by density waves, the density-wave theory predicts an angular rotation of the spiral pattern at a constant rate, while the gas and stars orbit circularly around the Galactic Center [GC] at a constant speed. Somewhere, at the corotation radius $R_{coro}$, the angular rotation of the gas and stars equals that of the angular rotation of the spiral pattern.

Determining objectively the spiral pattern's angular rotation $\Omega_{sp}$ has proven difficult. There is no readily observable chemical tracer able to do that measure directly. Hence many have resorted to a combination of properties and assumptions (some known, some hazy) in order to derive where the corotation radius is located.

Many theoreticians derived different values for corotation, depending on different methods and assumptions. Optical tracers (e.g., Cepheid stars) sometimes used to model the spiral pattern show a random distribution around the Sun in a disc's view from above (e.g., fig. 1 in Dambis et al 2015). When the arms are difficult to see in the disc's plane, then the error bars in the derived arm parameters are large.

**Before the corotation radius**, the angular speed of the stars and gas in the Sagittarius, the Scutum, and the Norma arms is *larger* than the angular speed of the spiral pattern, causing a shock when the circularly orbiting gas reaches and enters the spiral pattern from the *inward* side closest to the GC. Within the shocked lane, starforming regions are born, and these new stars will evolve across the arms and along their circular orbit around the GC; these aging stars will 'exit' on the other side of the arm. A sketch of this behavior can be seen in Martinez-Garcia et al (2009 – their fig.1) and in Shu (2016 – his fig. 8). This 'start' to 'exit' separation can be seen as a motion *across* a spiral arm (when viewed from above the disc) or as a dust-to-CO 'tracer offset' (when viewed from the Sun within the disk, tangentially along the galactic longitudes).

Earlier, we discovered a separation of about 315 parsecs between the dust/maser lane and the mid-arm CO tracer in the Sagittarius, Scutum, and Norma spiral arms, using observed tracer peaks along the galactic longitudes (inward from the Sun's orbit in the galaxy). In order to observe such a tracer separation outward from the Sun's orbit, in the Perseus arm, one needs a linear resolution of at least 100 parsecs (at the 3-sigma level) toward the Perseus arm, available only through the triangulating parallax method, which provides error bars in the measured distances at the level of about 1%.

Using precise parallax distances may test for the presence of a spatial offset of masers from the Perseus mid-arm.

Beyond the corotation distance, the density wave theory predicts that the speed of the orbiting gas and stars is *smaller* than that of the density wave spiral pattern, so that there is a shock when the spiral pattern reaches and overtakes the slower gas on the *outside* edge of the arm (the dust/maser lane is now located on the *outer* side of the arm, away from the direction of the GC, toward the anti-GC).

Here, we seek to find the radius $R_{coro}$ by looking at the tracer offset at each successive arm, starting from the GC and going out to the edge of the Galaxy. Thus when one can measure an offset among chemical tracers (dust versus CO gas, say), then one could check if the offset occurs inward or outward of the mid-arm.

## 1.1 Below the Sun's orbit.

Most recent determinations of the distance from the Sun to the GC indicate around 8.0 ± 0.2 kpc (Vallée 2017a).

Below the Sun's orbit, for the inner spiral arms, one can look at the arm tangent from the Sun, in order to get the galactic longitude where each tracer is observed at its maximum peak. Thus for the Sagittarius arm, its starforming regions (protostars, dust lane) are seen from the Sun to be at the inner arm side: the arm tangent for the dust lane is at galactic longitude $+48^{o}$, while the arm tangent for the CO mid-arm is at $+51^{o}$ (Table 1 in Vallée 2016). Its counterpart, the Carina arm, also shows the arm tangent for the dust lane to be at galactic longitude $-75^{o}$, while the arm tangent for the CO mid-arm is at $-79^{o}$ (Table 1 in Vallée 2016).

For all of the inner arms, one finds an offset between the dust lane and the main CO mid-arm (Table 4 and Fig. 3 in Vallée 2014) with the dust lane on the *inner* side of an arm toward the GC. The mean separation from the dust lane to the CO mid-arm is about 315 pc (Table 1 in Vallée 2016). Observationally, the galactic longitudes of each spiral arm tracer (dust, cold CO) show a reversal on both sides of the galactic longitude zero (Vallée 2016; Vallée 2017c).

1.2 Above the Sun's orbit.

Above the Sun's orbit, for the outer spiral arms, one cannot look at arm tangents from the Sun. However, one can look at the distance offsets of some tracers across a spiral arm, when measured through trigonometry.

Is the Perseus arm located inside or outside the corotation radius? Equivalently, for the Perseus arm, are the tracer offsets located on the *inner* edge or on the *outer* edge of the Perseus arm ? Observationally, where is this corotation distance $R_{coro}$ in the Milky Way disk?

1.3 Plan of this study.

Section 2 assembles recent observations of arm tracers having accurate distances, and compares them with a recent modeling of the spiral arm, in both space and velocity. Section 3 assembles and performs some statistics on the angular speed of the spiral pattern. Section 4 concludes on the likely location of the corotation radius in the Milky Way.

## 2. The Perseus arm, masers and galactic HII regions – tracer offsets

The use of trigonometrically derived accurate parallax distances from the Sun is important here, since their small error bars (near 1%) allow a better comparison with CO-derived models of spiral arms (themselves fitted to well-observed accurate galactic longitudes of the arm tangents below the Sun's position). Then we can look for an accurate offset in space between masers and models (such an offset near 0.3 kpc was found for inner spiral arms – see Vallée 2016).

Thus an observational accuracy of that order is needed for the distance to tracers. To get that required spatial accuracy, we culled from the literature the trigonometric parallax distance to many masers and their associated HII regions near the Perseus arm, in the second Galactic quadrant [GQ]. We cannot get that required spatial accuracy when using kinematically-derived or photometrically-derived distance estimates.

**Table 1** lists such galactic masers and HII regions and their recently determined trigonometrically-derived accurate parallax distances from the Sun. For examples, the HII region S235 near l=174$^o$ at 1.8 kpc from the Sun is well seen at the optical wavelengths (Dewangan & Ojha 2017). Similarly, the masers and HII regions NGC 7538, NGC281, IRAS 00420, WB 89-437 and S252 near the Perseus arm have a well-measured distances, as well as S269 and WB 89-437 near the Cygnus arm (Reid et al 2009). The starforming region W3(OH) at l=134$^o$ near 2.0 kpc kpc from the Sun was also well measured in 2001-2002 (Hachisuka et al 2006) and repeated in 2003-2004 (Xu et al 2006). Also, the giant HII region IC 1805 at l= 135$^o$ at 2.0 ± 0.2 kpc from the Sun and at a radial velocity of -47 km/s is well seen at optical wavelengths, around the Perseus spiral arm (Vallée et al 1979).

2.1 Comparisons with a recent arm model – a different offset for each different tracer

Various arm models have been proposed and compared (see Table 3 in Vallée 2017b). Here we employ the most recent spiral arm model, with an explanation for the GQ II and III in Vallée (2017b – Sect. 7) and for the inner GQ I and IV in Vallée (2017d – Sect. 3). Basically, this model fitted the CO 1-0 gas, peaking at the galactic longitude of the tangent to the spiral arm, as seen from the Sun (given in Table 5 of Vallée 2016).

It was then found that these CO longitudes (in mid-arm) differ by about 315 pc from the galactic longitudes where the dust lane and masers are seen tangentially (on the inside of the arm, toward the GC) – see Vallée (2016 – his fig.1 and table 1). This CO-fitted model was then extended close to the GC (Vallée 2017d) and outward (here).

Below, we compare the distances and velocities of the masers and HII regions to the CO-modelled Perseus arm, looking for distance offsets and velocity offsets.

**Figure 1** shows the sources in Table 1 toward the anti-GC (filled circles), as well as a recent spiral arm model (curves), as explained in Vallée (2017b – his fig.5a) and Vallée (2017d – his fig 3).

The model in Figure 1 employed the following characteristics: Sun to GC distance of 8.0 kpc (Vallée 2017a), mean arm pitch angle of -13.1$^o$ (Vallée 2017b), and logarithmic spiral shape (Vallée 2017c). The model fitted well the galactic longitudes of the arm tangents in the CO tracers for the Sagittarius, Scutum, and Norma arms, as observed from the Sun and cataloged in Tables 3 to 10 in Vallée (2016), hence the CO separation from the maser lane is well obtained by different observed tangencies for the two tracers.

For the Perseus arm, it can be seen in Figure 1 that most (about 85%) of masers/HII regions and in GQ II, with recently accurately determined distances, are located *inward* of the CO lane (middle of the arm) – at a mean offset of about 0.4 ± 0.1 kpc from the CO tracer. A roughly similar value is found in GQ III, but with much less data.

This mean offset, and its direction (inward), are similar to that found earlier for the other inner arms (Vallée 2014). The implication from the density wave theory of this maser offset from the cold CO mid-arm and inward arm location is that the Perseus arm lies inside the corotation radius.

In the GQ III, the Perseus spiral arm gets farther away from the Sun, and observational parallaxes get smaller and difficult to observe.

**Figure 2** shows the same sources from Table 1, and the radial velocities from a recent spiral arm model (Vallée 2017b – his figure 5b; Vallée 2017d – his fig 4). The CO tracer for the Perseus arm is shown in yellow.

The model spiral arm model in Figure 2 used the latest values for the orbital circular speed around the GC of 230 km/s (Vallée 2017a), and for the start of each spiral arm near the GC of 2.2 kpc (Vallée 2017b). This latest velocity model compares favorably with earlier published but incomplete models (Table 3 and Section 7 in Vallée 2017b).

For the Perseus arm, it can be seen in Figure 2 that most (about 85%) masers and HII regions in Galactic Quadrant II (at left) have a more negative velocity than the CO tracer of the Perseus arm – at a mean offset of about 9 ±2 km/s from the CO tracer. The mean velocity offset, and its direction (more negative), follow the prediction of the density wave theory.

In the third GQ, the Perseus spiral arm gets much distant, and the signal strengths get weaker. A roughly similar value is found in GQ III, but with much less accuracy.

2.2 Corotation beyond the Perseus arm - tracer offsets toward l=180$^{o}$

The observations shown in Figure 1, notably the location of the masers-HII regions being closer to the GC than the CO mid-arm, and the observations shown in Figure 2, notably the velocity of the masers and HII regions being more negative than that of the CO mid-arm in quadrant II, together make a strong suggestion that the angular speed of the stars and gas there is *larger* than the density wave's angular speed near the Perseus arm.

Therefore, the density wave's corotation radius must be beyond the Perseus arm - thus *beyond* a galactic radius $R_{coro}$ of 10.8 kpc (the Perseus arm distance at longitude 180$^o$ in Figure 1).

At that possible $R_{coro}$ value of 10.8 kpc, past the Perseus arm and closer than the Cygnus arm, the circular speed of gas and stars there (230 km/s – see Vallée 2017a) and the Perseus arm distance to the GC (10.8 kpc) give an angular velocity of gas and stars there of $\Omega_{gas} = V_{rot}/R = 21.3$ km/s/kpc.

If this galactic distance is at the 'corotation' point, then the density wave spiral pattern is at $\Omega_{sp} = 21.3$ km/s/kpc for the Milky Way. If not, then $R_{coro} > 10.8$ kpc, and there $\Omega_{sp} < 21.3$ km/s/kpc.

2.3 Corotation beyond the Perseus arm - tracer offsets in velocity at l=90$^o$

The observations shown in Figure 2, notably the velocity offsets of masers and HII regions with respect to the CO mid-arm, are similar to the 'streaming' predicted in the density-wave theory. Roberts (1975 – his Fig. 12) predicts an offset of around 30 km/s along a *circular* orbit.

Here in Fig. 2 we observe a *radial* velocity offset near 0 km/s at l=180$^o$, then near 15 km/s at l=140$^o$, near 25 km/s at l=115$^o$, and we can extrapolate it near 35 km/s at l=90$^o$. For that longitude of 90$^o$, the masers have a reduced orbital speed (230 – 35 = 195 km/s), at a galactic radius of 9.4 kpc, giving an $\Omega_{sp} = V_{maser}/R_{Perseus} < 20.7$ km/s/kpc. This corresponds to $R_{coro} > 11.1$ kpc

CO arm tracers employed in the model are the 'diffuse' CO emissions with a broad angular size (about 9 arcmin; Vallée 2016) coming from a large chunk of the spiral arm, but not the narrow CO cloud tracers harboring the maser emission from local starforming regions (and at a different radial velocity than the mid-arm's velocity).

These spatial offset and velocity offsets are definite, and show that the Perseus arm is definitely inside the co-rotation radius, since here the orbiting gas and stars must hit the inside of the arm to produce the inward spatial offset (Fig. 1) and the more negative velocity offset (Fig.2). This is a main result of this work, thanks to the precise parallax distances to masers near the Perseas arm.

To get a rough estimate of the location of the corotation radius, we turn to the recent literature.

## 3. Statistics

Observationally, the angular rotation for the gas and stars near the Sun at 8.0 kpc is $\Omega_{gas} = 28.8 \pm 1$ km/s/kpc, in the Milky Way Galaxy (table 2 in Vallée 2017a; Nagayama et al 2011). This value is much larger than the angular velocity of the corotation pattern $\Omega_{sp} < 21.3$ km/s/kpc.

Telescopes having made scans of intensity versus galactic longitudes have detected the dust lane and other tracers tangentially to many arms, but there was no tangents detected towards the 'Local Arm' near the Sun (table 3 in Vallée 2016).

3.1 Filtering

Some past reviews have pegged the angular rotation of the spiral arm's pattern over a wide range of values, from 17 to 30 km/s/kpc (Gerhard 2011) or even 12 to 30 km/s/kpc (Foster & Cooper 2010), yet each such prediction had a small published error bar of around 1 km/s/kpc. Thus the range of the predicted pattern speed $\Omega_{sp}$ is much larger than the mean quoted error.

The prediction is often obtained after a complex model is employed, along with some hypotheses. These values transform to a corotation radius estimate from 7.7 to 19 kpc (for $V_{gas}$ = 230 km/s).

**Table 2** shows a summary of recently predicted values for $R_{coro}$ , since 2010, as based on modeling of some observational data, excluding those with results having the corotation radius below 10.0 kpc (see below). It is expected that later predictions would have somewhat smaller error bars, hence our listing of predictions published since the year 2010.

Many such values were obtained using tracers whose distance estimates have a large error bar; distance estimates near 10% are obtained from various techniques, except the trigonometric parallax technique.

Large distance errors do not allow us to see a tracer (maser, say) being offset from the CO mid-arm by 0.3 kpc. Outward from the Sun's orbit, one needs a linear resolution of at least 100 parsecs toward the Perseus arm; only the parallax method provides error bars at the level of about 1%.

Foster & Brunt (2015 – Fig. 5a) investigated the photometric / spectroscopic / luminosity distance (with large error bars) to HII regions in the Perseus arm, noting conflicting distance estimates with parallactic distances (with small error bars) from the literature.

3.2 Statistics

Our results above (Sect. 2) indicate that the Perseus arm has a maser lane which is definitely offset from the main mid-arm CO, *thus* the Perseus arm is definitely inside the co-rotation radius, so then the $R_{coro}$ is definitely greater than 10.8 kpc. This definiteness is important, as it follows here that one can set aside all the predictions from the literature having smaller $R_{coro}$ values (based on non-trigonometric, non-parallax distances, and various assumptions).

Thus, one can perform standard statistics for all data predicting a corotation radius beyond 10.8 kpc. These basic statistics yield a mean $R_{coro}$ of 12.0 ±1.2 kpc and $\Omega_{sp}$ of 19.2 ±1.7 km/s/kpc.

3.3 Assumptions

This research assumes a global spiral pattern, with a single pattern speed. This assumption follows the work of many theoretical pioneers, reviewed already in Shu (2016). This implicitly assumes the usual 'quasi-stationary' galaxy-wide spiral density wave (global SDW), with a single spiral pattern speed. This assumption is backed by predictions of a spatial offset between the hot dust / maser density peak and the cold molecular / stellar spiral arm, as observed with telescopes, and of a mirror image of this offset as one crosses the Galactic Meridian (e.g., Vallée 2016). Over a whole galactic disk, the global SDW has a dust offset from the molecular arm, being inward below the corotation radius, and outward above the corotation. Inside corotation, the observations show a spatial separation between the CO mid-arm and the dust/maser lane, of about 315 pc, over the galactic disk, leading to a mirror image of this separation as one goes across the Galactic Meridian (Vallée, 2016).

Other theories assumed a local 'transient' spiral pattern. Thus, Baba et al (2016) employed a recurrent dynamic spiral theory (RDS), in which the gas does not flow through a spiral arm, but flows into the arm from both inner and outer sides (their Fig. 5). The RDS may have some local offsets, but not in a global systemic way over a whole galactic disk; thus the dynamic spiral theory denies the existence of a global tracer offset (e.g. Baba et al. 2016), but it can be "local". The transient-recurrent 'dynamic spiral' reconnection theory (Dobbs & Baba 2014, Sect. 2.2) has arms breaking and reconnecting, and a high pitch angle from $20^o$ to $40^o$ (Dobbs & Baba 2014, fig.10), unlike the mid $13^o$ pitch in the Milky Way. Also, Baba et al (2018) proposed disruptions in the Perseus arms, using Cepheid distances; these distance estimates are based on photometric period-luminosity techniques, not on trigonometric parallaxes, and have error bars too large for comparisons (near 10%; Vallée, 2017c). The RDS has no global, systemic radial dust offset from the cold molecular mid-arm (Sect.4 in Baba et al 2016).

The theory of stochastic, self-propagating star-formation (SSP) produces an irregular, flocculent arm (Dobbs & Baba 2014, sect. 2.5), not the regular pattern observed in the Milky Way.

The theory of tides between two nearby galaxies via kinetic density waves (KDW) predicts only two tidal arms, not four as seen in the Milky Way, and no spatial offset between tracers. The induced spiral arms will rotate slower than the galaxy rotates, but there is no global galaxy-wide offset of hot dust versus the cold molecular arm at most times (Dobbs et al 2010; Pettitt & Wadsley 2018).

This research also assumes a constant circular speed for the gas and stars, from a minimum galactic radius of 3 kpc going outward. This assumption has been verified observationally out to a galactic radius of 15 kpc (e.g., a review by Foster and Cooper 2010 – Fig. 2). Here we concentrate with the Perseus arm, around the galactic radius of 10 kpc (e.g., Foster and Brunt 2015 – Fig.5b), where the constant circular speed is verified.

## 4. Conclusion

The precise parallax distances to masers allowed the precise determination of a spatial offset of masers from the Perseus mid-arm, when compared to the latest arm model as previously fitted to the observed galactic longitudes of arm tangents (broad CO mid-arm).

That tracer offset is predicted only in the global SDW theory, thus giving it a precise advance over all other different theories (local RDS, SSP, KDW, etc.) for the Milky Way disk.

It also puts a precise minimum location for the corotation radius, and allows us to set aside all predictions to corotation radius below the Perseus arm.

Here, we performed a test of the tracer offsets in space associated with a spiral density wave, notably for masers and young HII regions close to the Perseus arm. We find here that the masers and young HII regions are located on the *inside* edge of the CO-traced arm (offset by about 0.4 kpc), and the ensuing corotation radius must be beyond this mid-arm, i.e. $R_{coro} > 10.8$ kpc from the GC. This implies that the angular rotation of the spiral pattern $\Omega_{sp} < 21.3$ km/s/kpc (see Sect. 2.2).

Next, we perform a test of the tracer offset in velocity, as a function of galactic longitude. We find that the masers and HII regions are located in the second quadrant at a more negative velocity (offset by about 9 km/s depending on longitude). This implies that the angular rotation of the spiral pattern $\Omega_{sp} < 20.7$ km/s/kpc, and $R_{coro} > 11.1$ kpc from the GC (see Sect. 2.3).

Together, the spatial offset and the velocity offset (inner arm, and negative velocity) as observed for the Perseus arm in GQ II follow a similar behavior found earlier for the inner spiral arms (Sagittarius, Scutum, Norma) located before the galactic corotation radius. These two offsets allow one to deduce that the Perseus arm is inside the corotation radius, a highlight of this paper. These offsets would have been seen in reverse, if the Perseus arm was beyond the galactic corotation radius (see Section 1).

In addition, we perform some statistics on the recently published values for the angular rotation rate of the density wave spiral pattern. Here we excluded the predictions with a $R_{coro}$ <10.8 kpc. We find a statistical mean $R_{coro}$ of about 12 kpc (when using a GC-Sun distance of 8.0 kpc), and a mean of $\Omega_{sp}$ of about 19 km/s/kpc (see Sect. 3).

These observational results should constrain future density-wave models of the Milky Way.

**Future works.** For $R_{coro}$ near 12 kpc and $\Omega_{sp}$ near 19 km/s/kpc, it appears that the inner 4:1 resonance is near 8.2 kpc (near the Sun) – see Hunt & Bovy (2018). This 4:1 resonance may be searched in the upcoming Gaia DR2 stellar velocity plots versus azimuthal, radial, longitude and latitude (Hunt & Bovy 2018; Kawata et al 2018; Ramos et al 2018).

The Lépine et al (2017) model, employing four trapped islands with horseshoe orbits for the gas near the corotation radius, might well work beyond the Perseus arm (nearer 12 kpc from the GC). Dias & Lépine (2005) followed backward the orbits of open clusters to deduced their place of birth and compared with the time displacement of their model arms, with many assumptions.

Tidal effects from the passages of the Sagittarius dwarf galaxy near the Milky Way may compete with the SDW and RDS models in interpreting the Gaia DR2 velocity data (Antoja et al 2018).

**Acknowledgements.** The figure production made use of the PGPLOT software at NRC Canada in Victoria. I thank an anonymous referee for useful, careful, insightful and historical suggestions.

Table 1. Sources toward the anti-GC (100$^o$ < l < 250$^o$), with a well-measured distance (since 2006).

| Name | Gal.long (o) | Gal. lat. (o) | Distance (kpc) See Note 1 | Syst. $V_{lsr}$ (km/s) | Reference |
|---|---|---|---|---|---|
| G100.37-3.57 | 100.4 | -3.6 | 3.46 ±0.20 | -37 | Choi et al (2014) |
| G108.20+0.58 | 108.2 | +0.6 | 4.41 ±0.90 | -49 | Choi et al (2014) |
| G108.47-2.81 | 108.5 | -2.8 | 3.24 ±0.10 | -54 | Choi et al (2014) |
| G108.59+0.49 | 108.6 | +0.5 | 2.47 ±0.22 | -52 | Choi et al (2014) |
| G110.19+2.47 | 110.2 | +2.5 | 3.18 ±0.90 | -63 | Chibueze et al (2014) |
| G111.23-1.23 | 111.2 | -1.2 | 3.33 ±1.23 | -53 | Choi et al (2014) |
| G111.25-0.77 | 111.3 | -0.8 | 3.34 ±0.27 | -43 | Choi et al (2014) |
| NGC 7538 | 111.5 | -0.8 | 2.65 ±0.12 | -57 | Choi et al (2014) |
| NGC 7538 | 111.5 | +0.8 | 2.65 ±0.05 | -57 | Reid et al (2009) |
| IRAS 00420 | 122.0 | -7.1 | 2.13 ±0.05 | -44 | Reid et al (2009) |
| NGC 281 | 123.1 | -6.3 | 2.82 ±0.05 | -31 | Reid et al (2009) |
| W3(OH) | 134.0 | +1.1 | 1.95 ±0.04 | -44 | Xu et al (2006) |
| W3(OH) | 134.0 | +1.1 | 2.04 ±0.07 | -49 | Hachisuka et al (2006) |
| W3(OH) | 134.0 | +1.1 | 1.95 ±0.04 | -45 | Reid et al (2009) |
| S Per | 134.6 | -2.2 | 2.42 ±0.11 | -39 | Choi et al (2014) |
| WB 89-437 | 135.3 | +2.8 | 5.99 ±0.22 | -72 | Reid et al (2009) |
| G160.14+3.15 | 160.1 | +3.2 | 4.09 ±0.09 | -18 | Reid et al (2014) |
| G168.06+00.82 | 168.1 | +0.8 | 5.00 ±0.5 | -28 | Hachisuka et al (2015) |
| IRAS05168+36 | 170.7 | -0.2 | 1.88 ±0.21 | -19 | Sakai et al (2012) |
| IRAS05168+36 | 170.7 | -0.2 | 2.02 ±0.13 | -19 | Sakai et al (2012) |
| Sharpless 235 | 174.0 | +3.4 | 1.8 ±0.2 | -18 | Dewangan & Ojha (2017) |
| G182.67-3.26 | 182.7 | -3.3 | 6.4±0.6 | -8 | Hachisuka et al (2015) |
| G183.72-3.66 | 183.7 | -3.7 | 1.59 ±0.03 | +3 | Choi et al (2014) |
| G188.79+1.03 | 188.8 | +1.0 | 2.02 ±0.35 | -5 | Reid et al (2014) |
| Sharpless 252 | 188.9 | +0.9 | 2.10 ±0.03 | +8 | Choi et al (2014) |
| Sharpless 252 | 189.0 | +0.9 | 2.10 ±0.05 | +11 | Reid et al (2009) |
| G192.60-0.04 | 192.6 | -0.0 | 1.52 ±0.09 | +6 | Choi et al (2014) |
| Sharpless 269 | 196.4 | -1.7 | 5.29 ±2.0 | +20 | Reid et al (2009) |
| Sharpless 269 | 196.4 | -1.7 | 5.28 ±0.23 | +19 | Hachisuka et al (2015) |
| G211.59+1.05 | 211.6 | +1.1 | 4.39 ±0.14 | +45 | Reid et al (2014) |
| G229.57+0.15 | 229.6 | +0.2 | 4.59 ±0.27 | +47 | Choi et al (2014) |
| G236.81+1.98 | 236.8 | +2.0 | 3.07 ±0.27 | +43 | Choi et al (2014) |
| G240.31+0.07 | 240.3 | +0.1 | 5.32 ±0.49 | +67 | Choi et al (2014) |

Note 1: Excluding nearby sources, identified as located in the ‘local arm’ or spur - see Reid et al (2014). When needed, the published parallax (p, in mas) was converted to a distance (D, in kpc) through the equation D = 1/p.

**Table 2 – Published recent (since 2010) angular rotation of the spiral pattern speed $\Omega_{sp}$ and deduced corotation radius $R_{coro}$ (10 kpc and above).**

| $\Omega_{sp}$ (km/s/kpc) | **observed data** used in model | $R_{coro}$ (kpc); Note 1 | **References** |
|---|---|---|---|
| 16 ±1 | terminal HI velocity | 14.3 | Sect.3.2.2 in Foster & Cooper (2010) |
| 17 ±0.6 | stellar kinematics | 13.5 | Sect.12 in Antoja et al (2011) |
| 18.6 ±0.3 | radial velocity of stars | 12.4 | Sect. 3.2 in Siebert et al (2012) |
| 19.3 ±0.8 | peculiar vel. V=0 | 11.9 | Sect. 4.2 in Sakai et al (2015); see Note 2 |
| 20 ±5 | HI and CO data | 11.5 | Table 2 in Koda et al (2016) |
| 20 ±2 | CO data | 11.5 | Sect. 5 in Pettitt et al (2014) |
| 20.3 ±0.5 | early-type stars | 11.3 | Sect. 3.2 in Silva & Napiwotski (2013) |
| < 20.7 | maser veloc. Offset | > 11.1 | This paper (Section 2.3) |
| < 21.3 | maser offset in space | > 10.8 | This paper (Section 2.2) |
| | | | |
| 23 ±0.5 | open star clusters | 10.0 | Table 4 in Junqueira et al (2015) |
| 23 ±0.5 | HI gas flow | 10.0 | Sect. 2.2.3 in Li et al (2016) |

Note 1: we deduce $R_{coro}$ from the relation $R_{coro}$ **.** $\boldsymbol{\Omega_{sp}}$ **=** $\mathbf{V_{gas}}$, with $V_{gas}$ = 230 km/s (Vallée 2017a).

Note 2: their $R_{coro}$ =12.4 kpc value for their $R_{sun}$ = 8.33 is corrected here for our 8.0 kpc value (giving 11.9), and then the $\boldsymbol{\Omega_{sp}}$ value is deduced from our $V_{gas}$ value above.

**Figure captions.**

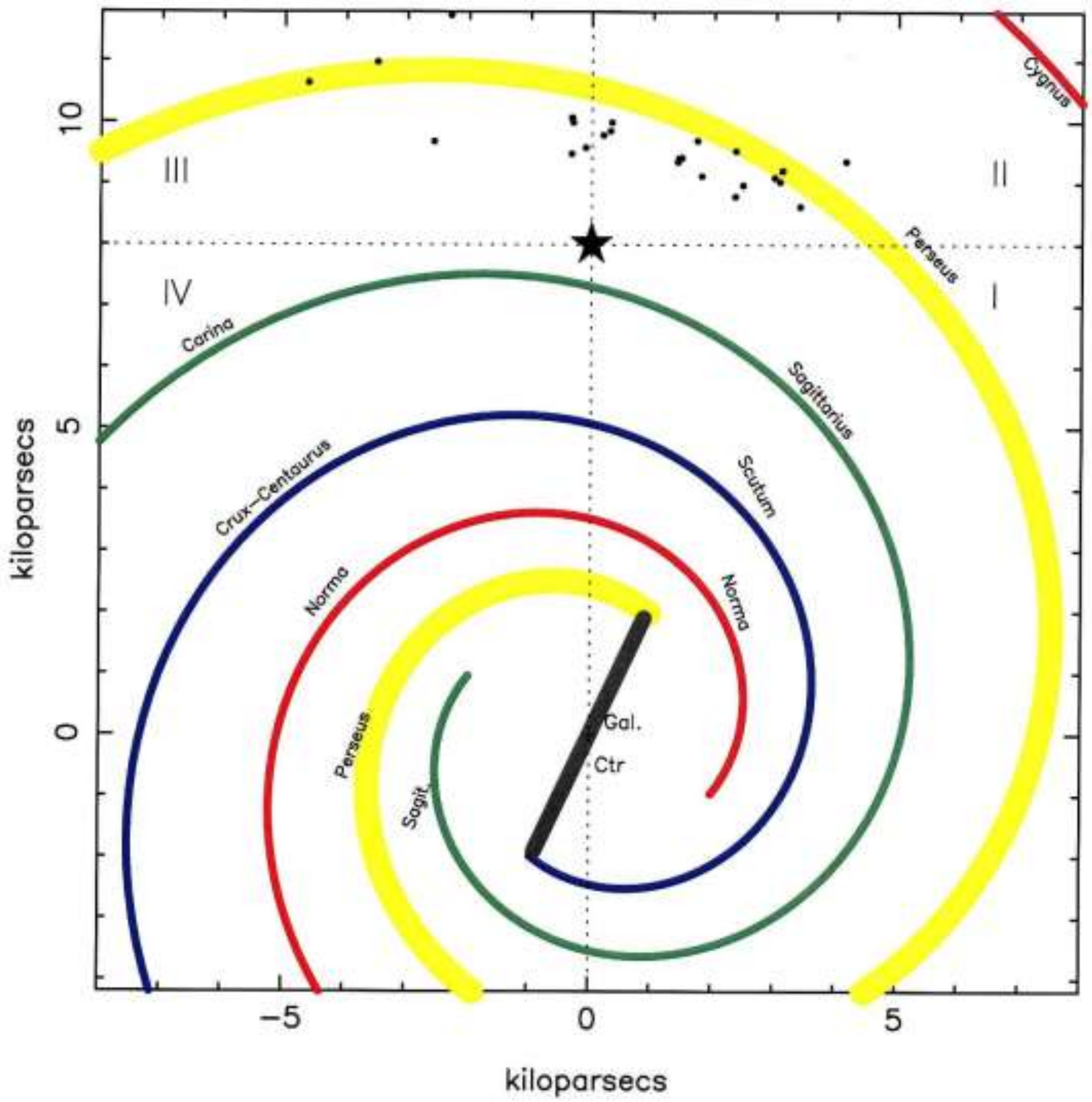

Figure 1. A view of the Milky Way's spiral arms (each arm in a different color), seen from above the galactic plane. The Perseus arm is in yellow. The Sun is sketched as a star at 8.0 kpc from the Galactic Center (at 0,0). Masers and HII regions with accurate distances near the Perseus arm are shown as filled circles. Galactic quadrants are shown: I (right bottom), II (right top), III (left top), and IV (left bottom).

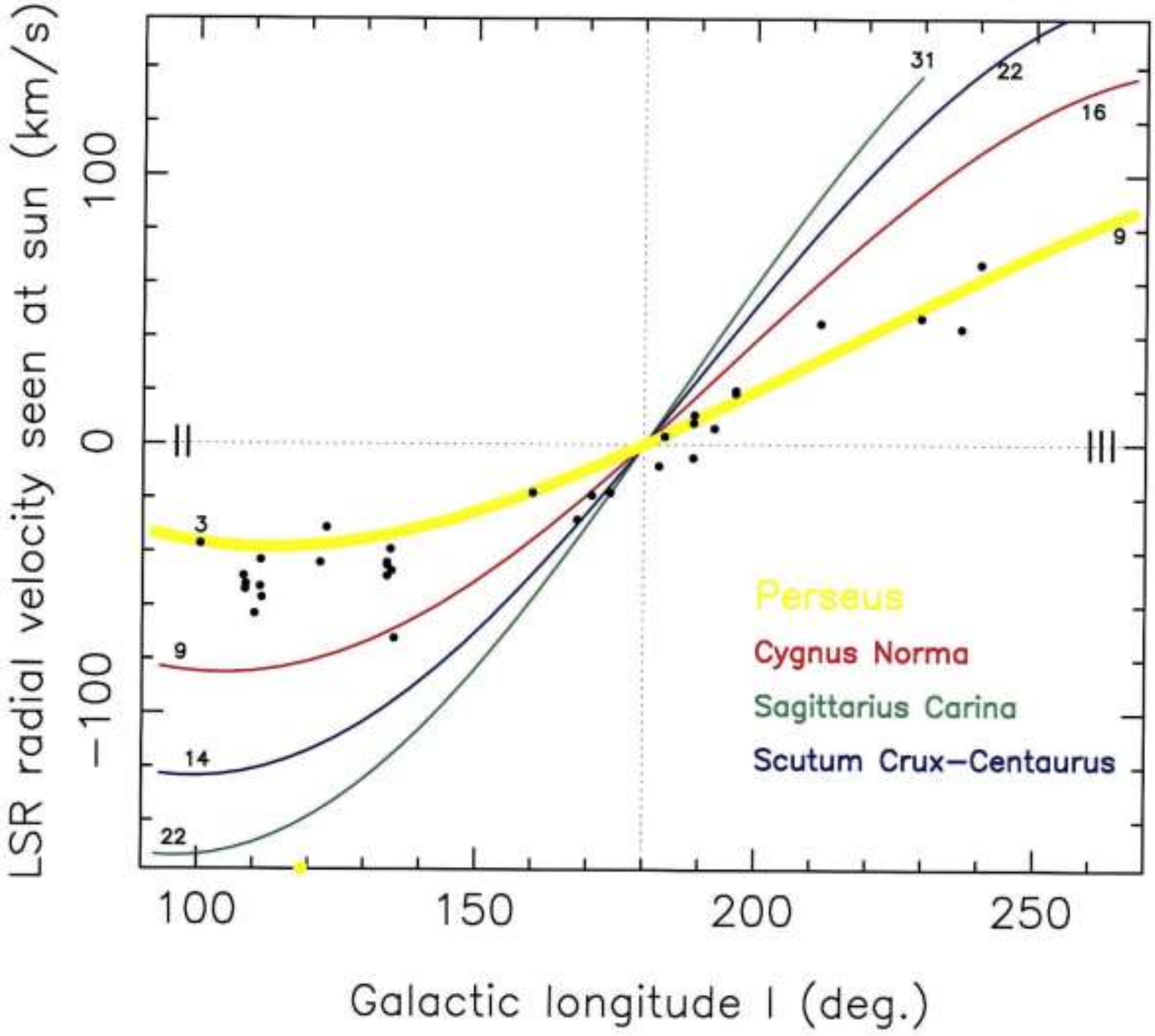

Figure 2. A view of the Milky Way's spiral arms (same color code as in previous figure) in radial velocity (km/s) versus galactic longitude (°), as seen from the Sun. The number indicated along a part of an arm (at bottom left or upper right) indicates the arm's rough distance from the Sun. Masers and HII regions with an accurate distance near the Perseus arm are shown as filled circles. Two galactic quadrants are shown: II (left), and III (right).